\documentclass[10pt,letterpaper]{article}
\usepackage[top=0.85in,left=2.75in,footskip=0.75in]{geometry}

\usepackage{changepage}

\usepackage[utf8]{inputenc}

\usepackage{textcomp,marvosym}

\usepackage{fixltx2e}

\usepackage{amsmath,amssymb}

\usepackage{cite}

\usepackage{nameref,hyperref}

\usepackage{microtype}
\DisableLigatures[f]{encoding = *, family = * }

\usepackage{rotating}

\raggedright
\usepackage[aboveskip=1pt,labelfont=bf,labelsep=period,justification=raggedright,singlelinecheck=off]{caption}

\makeatletter
\renewcommand{\@biblabel}[1]{\quad#1.}
\makeatother

\date{}

\newcommand{\q}{\phantom0}

\usepackage{color}

\usepackage{booktabs}

\begin{document}
\vspace*{0.35in}

\begin{flushleft}
{\Large
\textbf\newline{Indexing arbitrary-length $k$-mers in sequencing reads}
}
\newline
\\
Tomasz Kowalski\textsuperscript{1},
Szymon Grabowski\textsuperscript{1},
Sebastian Deorowicz\textsuperscript{2},
\\
\bf{1} Institute of Applied Computer Science, Lodz University of Technology, Al.\ Politechniki 11, 90-924 {\L}\'{o}d\'{z}, Poland
\\
\bf{2} Institute of Informatics, Silesian University of Technology, Akademicka 16, 44-100 Gliwice, Poland
\\

%
%





* E-mail: Corresponding sebastian.deorowicz@polsl.pl
\end{flushleft}
\section*{Abstract}
We propose a lightweight data structure for indexing and 
querying collections of NGS reads data in main memory. 
The data structure supports the interface proposed in the 
pioneering work by Philippe et al.~\cite{PSLKCR2011} 
for counting and locating $k$-mers in sequencing reads.
Our solution, PgSA (pseudogenome suffix array), 
based on finding overlapping reads, is competitive to the 
existing algorithms in the space use, query times, or both.
The main applications of our index include variant calling, 
error correction and analysis of reads from RNA-seq experiments.


\section*{Introduction}
The genome sequencing costs dropped recently to 
less than 5
thousand U.S.\ dollars per human genome with about 30-fold coverage~\cite{NHGRI2015}.
Using the recent (and expensive) Illumina HiSeq X Ten system~\cite{Hay2014}, 
it may be even possible to reduce this cost to about 1 thousand dollars 
(or somewhat more) on a long run.
The scale of the largest sequencing projects is amazing, e.g., the Million Veteran Program~\cite{MVP2015} aims at sequencing 1M human genomes.
Needless to say, all this results in enormous amounts of sequencing data.

These data have to be processed in some way. 
Usually, they are mapped onto reference genomes and then 
variant calling algorithms are used to identify the mutations present in sequenced genomes.
Since the mapping requires 
fast search over 
reference genomes, a lot of indexing structures for genomes were adopted or invented.
The obvious candidates were 
the suffix tree and the suffix array~\cite{Gus1997},
but their space requirements were often prohibitive,
especially in the beginning of the 21st century.
The situation changed with the advent of much more compact (compressed) index data structures.
The most widely used in the read aligning software is the family of FM-indexes~\cite{FM2000}, 
employed 
by the popular Bowtie~\cite{LS2012}, BWA~\cite{LD2009} and many other mappers.
Modern computers are more powerful, hence 
nowadays using a suffix array is often not
a problem, especially if the array is sparsified 
(i.e., only a fraction of indexes is represented explicitly)~\cite{KU1996}.
One of the recent successful examples is the MuGI multi-genome index~\cite{DDG2014}, 
allowing to index 1092 human genomes in 
less than 10\,GB of memory.

As said above, a lot was done in the area of genome indexing, but very little for the 
other standard component of the input, i.e., sequencing reads.
The main reason is that when the reads are simply mapped onto a 
reference genome,
indexing them is pointless.
In many situations, however, the reads are processed in some way before 
(or instead of) mapping.
The most obvious case is read correction, which makes the mapping (or de novo assembling) easier and yields better final results, i.e., better determination of mutations.
There are a number of read correctors, e.g., Quake~\cite{KSS2010}, RACER~\cite{IM2013}, BLESS~\cite{HWCMH2014}, Fiona~\cite{SWHDNRR2014}; see the recent survey~\cite{MI2014} for more examples.
Sometimes the paired reads are joined if they overlap, with benefits in the mapping quality~\cite{ZKFS2014}.
In some other applications, e.g., in metagenomic studies, the goal is to assign reads to species (to identify which organisms can be found in the analyzed probe), and the reads are not mapped at all~\cite{AHGLGA2013,WS2014,BC2012}.

In such cases no reference sequence is used (or it is used only implicitly) and all the available knowledge can be retrieved only from the reads.
The simplest approach is to calculate the statistics of $k$-mers (i.e., all $k$-symbol long substrings of reads), but some programs use more sophisticated knowledge.
Therefore, the necessity of indexing reads was identified recently~\cite{PSLKCR2011}.
Philippe et al. defined therein the index supporting the following queries.
Given a query string $f$:
\begin{itemize}
  \item[$Q1$] In which reads does $f$ occur?
  \item[$Q2$] In how many reads does $f$ occur?
  \item[$Q3$] What are the occurrence positions of $f$?
  \item[$Q4$] What is the number of occurrences of $f$?
  \item[$Q5$] In which reads does $f$ occur only once?
  \item[$Q6$] In how many reads does $f$ occur only once?
  \item[$Q7$] What are the occurrence positions of $f$ in the reads where it occurs only once?
\end{itemize}

There are two ways in which $f$ can be given in those queries, 
which may lead to different time complexities and actual timing results.
In one, $f$ is given as a sequence of DNA symbols.
In the other, $f$ is represented as a read ID followed with the start position 
of $f$ in this read (and optionally, $f$'s length, if it is not fixed).

There are a number of potential applications of this index.
Philippe et al.~\cite{PSLKCR2011} described the following.
The queries $Q1$ and $Q2$ can be used for mutation (both SNPs and short indels) detection.
The query $Q2$ can be used to calculate a ``local coverage'' of a $k$-mer, i.e., the number of reads sharing it.
This was used in the work~\cite{PSCR2013} for calculation of ``support profile'' of each $k$-mer in a large package for analyzing reads from RNA-seq experiments.
One more potential usage of index queries $Q3$ and $Q4$ can be in clustering and assembly without a reference genome.

One of the successful techniques in read correctors, e.g., BLESS, RACER, is to preprocess the reads to 
collect the $k$-mer frequencies
(i.e., allow to answer the $Q4$ queries), 
which can be obtained with specialized software~\cite{RLC2013,MK2011,DDG2013}.
In some other tools, like Fiona~\cite{SWHDNRR2014}, Shrec~\cite{SSPSS2009}, HybridShrec~\cite{S2010}, it is necessary also to obtain the list of reads containing the $k$-mer (i.e., they need $Q1$ queries).
The solution used in Fiona is to construct the generalized suffix array, i.e., suffix array containing all suffixes from all reads.
Unfortunately, this implies huge memory requirements, e.g., for reads of human sequencing with 10-fold coverage, the memory occupation is 224\,GB.

Currently, only a few indexing structures supporting the mentioned list of queries are known.
Historically, the first one is Gk arrays (GkA)~\cite{PSLKCR2011}.
This scheme 
works for a single length of $f$ only 
(set at construction time).
The main GkA idea is to order lexicographically all substrings of 
length $k = |f|$ of the reads.
Let us denote the cardinality of the reads collection with $q$.
Assume that the reads are of equal length $m$.
As the number of reads substrings is $q(m - k + 1)$, 
the binary search for sequences with a given $k$-long prefix, 
like in a suffix array~\cite{MM1993},
has time complexity of $O(k \log((m - k + 1)q))$. 
In the following we use the symbol $n = q(m - k + 1)$ to simplify notation.
This operation answers query $Q4$ with $f$ given as a sequence of symbols.
If, however, the query 
position 
is given, then $Q4$ is handled in constant time.
GkA is based on three arrays: one for storing the start position of each $k$-mer, 
one inverted array telling the lexicographic rank of a $k$-mer given its position 
in a read, 
and finally an array associating to a $k$-mer's rank its number of occurrences.
The proposed data structure was found to be both more memory efficient 
and (in most cases) faster than two alternatives, 
a hash table and a suffix array augmented with some helper tables.

V{\"a}lim{\"a}ki and Rivals~\cite{VR2013} proposed a 
compressed variant of Gk arrays, 
based on the compressed suffix array (CSA)~\cite{GGV2003}.
Their index, CGkA, reduces the size of its predecessor by about 40\% to 90\%,
handling most queries 
with similar time complexity.
Like GkA, this solution also supports a single value of~$k$.

The index presented in this paper is based on two ideas: 
building a pseudogenome by finding overlapping reads in the collection, 
and using the sparse suffix array~\cite{KU1996} as the search engine in 
the resulting sequence.
We performed a number of experiments to compare the proposed PgSA (pseudogenome suffix array) and the existing GkA and CGkA indexes for 
the supported queries.
Then, to see how PgSA would work in a real environment, we replaced the GkA in CRAC~\cite{PSCR2013} by our index 
to check its overall memory consumption and processing time.


\section*{Materials and Methods}
We assume that the input alphabet contains 4 (\texttt{ACGT}) or 5 symbols
(\texttt{ACGTN}). 
The actual number of symbols in the input data implies some design choices 
in the internal representation of our index.
By a {\em pseudogenome} we mean a sequence obtained by concatenation of 
all (possibly reverse-complemented) reads with overlaps.
More formally, let us have a read array $\mathcal{R} = [R_1, \ldots, R_q]$, 
where $|R_i| = m$ for all $i \in \{1, \ldots, q\}$.
A pseudogenome is a sequence $PG[1 \dots p]$ for which 
\begin{itemize}
\item there exists a sequence $j_1, j_2, \ldots, j_q$ such that 
$j_1 = 1$, $j_{i+1} - j_i \in \{0, 1, \ldots, m\}$ for all $i \in \{2, \ldots, q\}$ 
and $j_q = p - m + 1$, 
\item for each $j_i$ we have $PG[j_i \ldots j_i+m-1] = R_{u_i}$ or 
$PG[j_i \ldots j_i+m-1] = rc(R_{u_i})$, where $rc(\cdot)$ is the reverse-complement 
operation on a DNA sequence, 
\item $[u_1, u_2, \ldots u_q]$ is a permutation of $\{1, 2, ..., q\}$.
\end{itemize}
We attempt to minimize the pseudogenome length $p$.
In further considerations we usually deal with the permuted 
read array, hence we define it as $\mathcal{R'} = [R_{u_1}, \ldots, R_{u_q}]$, 
where the indices $u_i$ are described just above.

While a sequence approximating a {\em real} genome may be obtained 
by a de novo assembly procedure, we refrain from it because of two reasons.
First, our procedure is lightweight, at least in conceptual and programming sense, 
while the problem of de novo assembly is known to be hard.
Second, removing sequencing errors during the assembly is obviously beneficial 
for the output accuracy, but we aim at indexing original reads, 
and mapping the reads onto a ``corrected'' genomic sequence would 
complicate the index representation and would possibly be detrimental 
to query handling effectiveness.

Note that the minimal pseudogenome problem, without allowing the 
reverse-complement operations on the reads,
is known in string matching 
literature under the name of the shortest common superstring (SCS) problem.
SCS is NP-hard, as shown by Maier and Storer~\cite{MS1977}.

\begin{figure}
\centering\includegraphics[width=\textwidth]{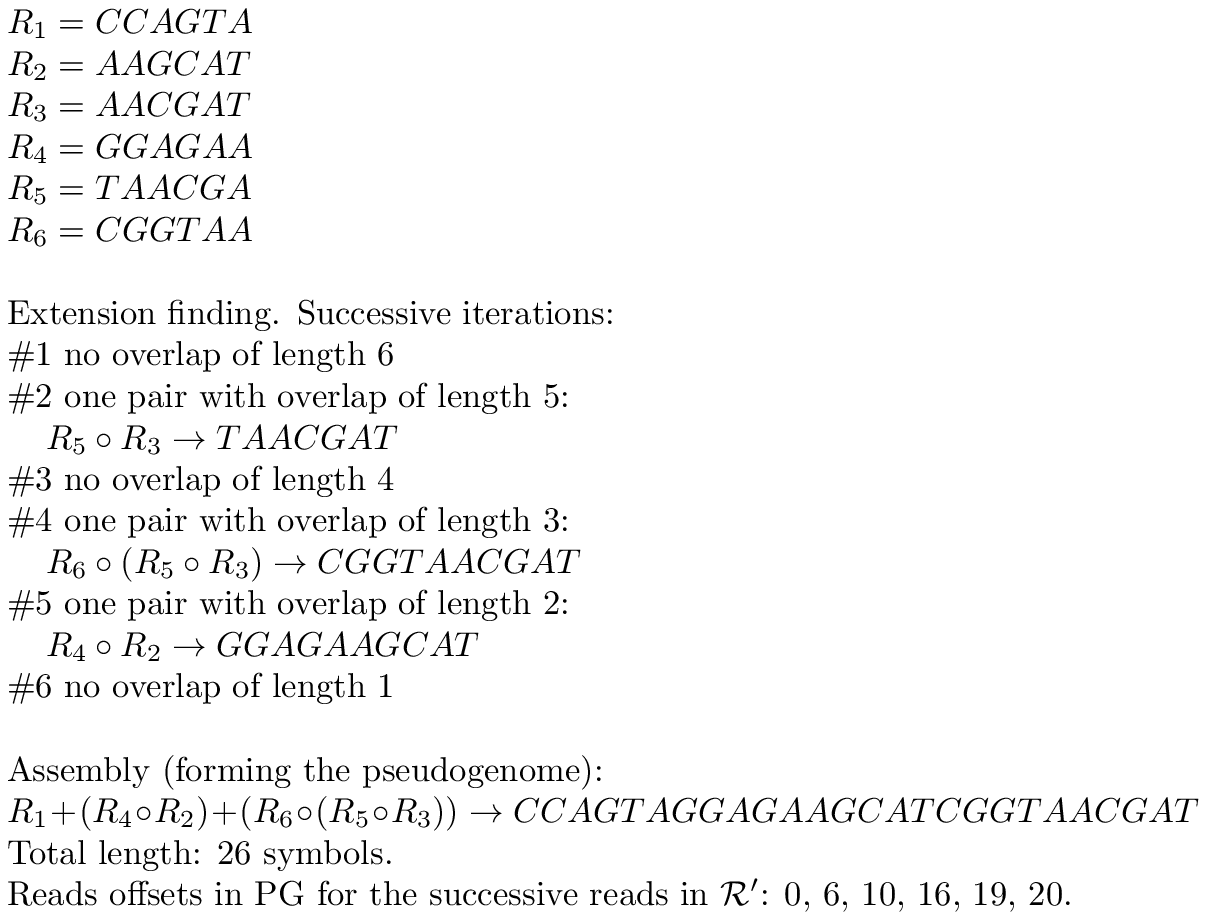}
\caption{Pseudogenome generation example.
The input read collection $\mathcal{R}$ contains 6 reads of length 6. 
We use two symbols: $+$ and $\circ$. 
$S + T$ is a plain concatenation of strings $S$ and $T$.
$S \circ T$ denotes a concatenation of strings $S$ and $T$ with a 
non-zero overlap of maximal length.}
\label{fig:pg_generation}
\end{figure}

The pseudogenome is generated in the following way. 
For each possible positive overlap length, $\mathit{ol} \in \{m, m-1, \ldots, 1\}$, 
considered in descending order, 
we look for reads that 
overlap any other reads. 
Yet, one read may be 
a successor of only one other read 
and also one read is disallowed to be 
followed (directly overlapped) by more than one read.
(Note that duplicate reads correspond to having 
$ol = m$.)
If some reads are left (i.e., are not 
successors
to any other read), 
they are attached at the end of the pseudogenome.
The construction worst-case time complexity is
$O(q m^2 \log q)$, where the logarithmic factor
comes from the balanced binary search tree based 
implementation of the C++ multiset container.
Note that $q m = \Theta(n)$ under the realistic assumption that 
$m-k+1 = \Theta(m)$, and then we can simplify the complexity formula to 
$O(n m \log q)$.
Finally, we point out that the practical performance of this 
algorithm is much better than
the worst case time suggests, due to many long overlaps in real data 
with large coverage.
Fig.~\ref{fig:pg_generation} illustrates.

Note that our current pseudogenome implementation does not handle reverse-complemented reads.
Yet, our preliminary experiments with adding reverse-complemented reads to the generated sequence 
resulted in rather moderate improvement in the pseudogenome length (e.g., shorter by about 15\%), 
while handling the queries requires significant changes in the used data structures (and possibly 
more space needed for them).
For this reason, we leave this harder problem version as a future work.

We note that this procedure is only a heuristic and does not guarantee 
to produce an optimal (shortest possible) pseudogenome.
To see this, consider an example of three reads:
$R_1 = CCAGTA$, $R_2 = AAGCAT$ and $R_3 = AACGAT$.
According to the presented algorithm, we obtain the assembly
$(R_1 \circ R_2) +  R_3 \rightarrow CAATGATAA$ of length 9.
Yet, the assembly $(R_1 \circ R_3) \circ R_2 \rightarrow CAATAATG$ 
produces a sequence of length 8.

The actual pseudogenome representation depends on 
the given data
(number of reads, read length etc.). 
In general it contains the $PG$ string and the read array $\mathcal{R}^\text{PG}$ consisting of either 9- or 13-byte records. 
Consecutive records correspond to consecutive reads in the pseudogenome 
and contain the following fields:
\begin{itemize}
\item read offset in the pseudogenome (4 or 8 bytes,
depending on the pseudogenome length),
\item flag data occupying 1 byte (duplicate $k$-mer flag, occurrence flag, single-occurrence flag, to be described later; several bits of this byte 
are not used),
\item read index in the original read array $\mathcal{R}$ (4 bytes).
\end{itemize}


Over the pseudogenome a search structure is built.
Our basic solution is based on the classic suffix array (SA)~\cite{MM1993}, 
as a simple and fast full-text index.
The $\mathit{SA}^\text{PG}$ elements require from 4 to 6 bytes.
One element, associated with one pseudogenome suffix, 
stores the following fields:
\begin{itemize}
\item a read array index of the furthest read (of $\mathcal{R}^\text{PG}$) containing starting symbols 
of the given suffix (3 or 4 bytes, depending on the number of reads 
in the collection),
\item start position of the suffix with regard to the beginning of the read 
(1 or 2 bytes, depending on the read length).
\end{itemize}


In order to 
access 
a suffix one has to obtain from the read array $\mathcal{R}^\text{PG}$
the offset of the specified read and add an offset of the suffix. 
Such organization enables straightforward identification of reads 
containing the sought prefix of the suffix.

Packing DNA symbols into bytes is a standard idea in compact data structures.
We adopt this solution for the pseudogenome, in order to reduce the space use, minimize the rate of cache misses during searches and boost string comparisons 
(due to a lesser number of compared bytes on average).
When the alphabet contains 4 symbols 
we handle 
the following compaction variants: $(i)$ 2, 3 or 4 symbols per byte, 
$(ii)$ 5 or 6 symbols per 2-byte unit (``short'').
For the 5-symbol alphabet 
we pack 
either $(i)$ 2 or 3 symbols per byte,
or $(ii)$ 4, 5 or 6 symbols per 2-byte unit.

Apart from the standard variant, we also implement a 
sparse suffix array (SpaSA)~\cite{KU1996}, 
which samples the suffixes in regular distances from the SA.
The distances between sampled suffixes are specified by input parameter $s$. 
More precisely, if the pseudogenome is represented with 
$PG[1 \ldots p]$ (w.l.o.g. assume that $s$ divides $p$), the SpaSA index contains $p/s$ suffix offsets: 
$s, 2s, \ldots, p$.
The data stored for a sampled suffix are like described above, 
plus $s-1$ preceding symbols, in packed form.
We set the $s \leq 6$ limitation.
Storing these $s-1$ symbols allows not to access the pseudogenome sequence 
during a scan over the suffix array (cf.~the $Q3$ query, described later) 
and is thus cache friendly.
To make the current implementation easier and faster (due to less conditions necessary to check in the search procedure) the sparsity of the suffix array determines the packing of symbols, e.g., $s=5$ means that 5 symbols are packed into 2-byte unit.
Note that the $s-1$ packed symbols require up to 2 bytes, 
hence the whole element for a suffix requires from 5 to 8~bytes.

For small values of $k$ it is feasible to precompute all answers for the 
counting queries ($Q2$, $Q4$, and $Q6$).
We assume the query is over the 4-symbol alphabet (\texttt{ACGT}).
When the pseudogenome is small (up to 300\,Mbases) we cache the answers 
for all $k \leq 10$, and for larges pseudogenomes for all $k \leq 11$.
The $Q2$ and $Q6$ results occupy 4~bytes each and $Q4$ results 8 bytes. (Handling $Q4$ needs more space since $f$ may appear in a single read several times.)

We note that the queries $Q2$, $Q4$, and $Q6$ are related. 
For example, the number of reads in which string $f$ occurs only once ($Q6$) 
is often not much smaller than the total number of occurrences of $f$ ($Q4$), 
and sometimes these values may be even equal; 
the equality of $Q4$ and $Q6$ also implies the same 
value of $Q2$.
We make use of this fact and store answers also for {\em some} longer $k$-mers: 
up to $k = 12$ using 2-byte units and single bytes for $k = 13$.
The precomputed answers are stored only if $Q2 = Q4 = Q6$, 
and $Q2$ less than $2^{16} - 1$ or $2^8 - 1$, depending on the used variant.
The opposite case is signaled on the 1- or 2-byte field with an unused value.

\begin{table}
\begin{adjustwidth}{-1.25in}{0in}
\caption{Worst-case time complexities. To save space, the $O(.)$ symbols 
around each formula were omitted.
$|Q3'|$ equals $|Q3|$ plus the number of visited SA$^\text{PG}$ locations.
Note that $n = q(m - k + 1)$.
}
\begin{tabular}{lccccc}\toprule
query & Gk (pos) & CGk (pos) & Gk (seq) & CGk (seq) & PgSA (pos/seq) \\
\midrule
Q1  & $|Q3|$ & $|Q1|\log\log n$ & $k\log n + |Q3|$ & $k\log\sigma + \newline 
       \operatorname{polylog} n + |Q1|\log\log n$ & $k\log p + |Q3'|$ \\
Q2  & $|Q3|$ & $\log\log n$ & $k\log n + |Q3|$ & $k\log\sigma + \newline 
       \operatorname{polylog} n$ & $k\log p + |Q3'|$ \\
Q3  & $|Q3|$ & $|Q3|\log\log n$ & $k\log n + |Q3|$ & $k\log\sigma + \newline 
       \operatorname{polylog} n + |Q3|\log\log n$ & $k\log p + |Q3'|$ \\
Q4  & $1$ & $\log\log n$ & $k\log n$ & $k\log\sigma + \newline 
       \operatorname{polylog} n$ & $k\log p + |Q3'|$ \\ 
Q5  & $|Q3|$ & $|Q5|\log\log n$ & $k\log n + |Q3|$ & $k\log\sigma + \newline 
       \operatorname{polylog} n + |Q5|\log\log n$ & $k\log p + |Q3'|$ \\
Q6  & $|Q3|$ & $\log\log n$ & $k\log n + |Q3|$ & $k\log\sigma + \newline 
       \operatorname{polylog} n$ & $k\log p + |Q3'|$ \\
Q7  & $|Q3|$ & $|Q7|\log\log n$ & $k\log n + |Q3|$ & $k\log\sigma + \newline 
       \operatorname{polylog} n + |Q7|\log\log n$ & $k\log p + |Q3'|$ \\
\bottomrule
\end{tabular}
\label{table:complexities}
\end{adjustwidth}
\end{table}

Table~\ref{table:complexities} compares the worst-case time 
complexities for the queries $Q1$--$Q7$ of the existing algorithms.
We use the notation $|Qx|$ to represent the number of occurrences reported by query $Qx$ (for $x=1,3,5,7$).
In the following paragraphs we describe how the seven queries are 
performed in an order dictated by exposition clarity.

\paragraph{$Q3$}
We binary search the suffix array $\mathit{SA}^\text{PG}$ for the string $f$, 
and for each potential match in the found range, 
pointing to some position in the pseudogenome $PG$ 
(represented as a pair: read ID in the read array $\mathcal{R}^\text{PG}$ and the suffix offset 
with regard to the beginning of the read), 
we check in how many (0 or more) reads $f$ really occurs.
To this end, we check if the suffix offset shifted by $k$ symbols 
does not exceed the read length $m$.
If this is the case, we add its position to the output list, 
otherwise we terminate.
Then, we scan over the read array $\mathcal{R}^\text{PG}$ backward,
adding a position as long as the suffix offset plus $k$ 
still does not exceed $m$.

\paragraph{$Q4$} We follow the procedure for $Q3$, but simply count the matches. 

\paragraph{$Q1$} This query is related to $Q3$, but requires filtering, 
as $f$ may occur in a read more than once.
To this end, ``occurrence flags'' (stored in flag fields of $\mathcal{R}^\text{PG}$) are used.
Initially, all these flags are set to false.
During the iteration over reads (like in the $Q3$ query) only non-visited yet reads
are added to the output list and for each new read the corresponding flag is set to true.
The flag locations are also put on a stack, to remove them in $O(|Q1|)$ time
at the end, leaving all ``occurrence flags'' set to false in $\mathcal{R}^\text{PG}$.
In general $|Q1| \leq |Q3|$, but since the equality often holds, we implemented some optimization.
The array $\mathcal{R}^\text{PG}$ stores ``repetitive read flag'' for each read.
This flag is true if the read contains at least one 11-mer at least twice.
When we process the reads answering the $Q1$ query we verify the flag. 
If it is false we are sure that no $f$ (of length at least 11) can appear in the read more than one time.

\paragraph{$Q2$} This query is to $Q1$ exactly like $Q4$ to $Q3$.

\paragraph{$Q5$} Again, this query is related to $Q3$, with extra filtration needed.
Now ``single-occurrence'' flags in $\mathcal{R}^\text{PG}$ are used.
The one-visit only mechanism for reads and unsetting the flags with aid of a
stack is identical as in $Q1$.
The operations on the stack take $O(|Q5|)$ time, where $|Q5| \leq |Q3|$.
Also here the ``repetitive read'' flags are used as a helpful heuristic.

\paragraph{$Q6$} This query is to $Q5$ exactly like $Q4$ to $Q3$, or $Q2$ to $Q1$.

\paragraph{$Q7$} We follow the procedure for $Q5$, only with replacing read IDs 
with the match positions.

\section*{Results}

We ran experiments to confirm validity of our algorithm in practice.
The testbed machine was equipped with an Intel i7 4930K 3.4\,GHz CPU 
and 64\,GB of RAM (DDR3-1600, CL11), running 
Linux 3.13.0-43-generic x86\_64 (Ubuntu 14.04.1 LTS).
Table~\ref{table:datasets} presents the datasets used in the tests.
These datasets are taken from:
\begin{itemize}
\item E.\ coli (11.5M reads of 151\,bp) --- \url{ftp://webdata:webdata@ussd-ftp.illumina.com/Data/SequencingRuns/}
\url{MG1655/MiSeq_Ecoli_MG1655_110721_PF_R1.fastq.gz},
\url{ftp://webdata:webdata@ussd-ftp.illumina.com/Data/SequencingRuns/}
\url{MG1655/MiSeq_Ecoli_MG1655_110721_PF_R2.fastq.gz},
this dataset was used in the CGkA paper~\cite{VR2013},
\item GRCh37 (42.4M reads of 75\,bp; no N symbols in the data) --- 
\url{http://crac.gforge.inria.fr/index.php?id=genomes-reads}, this dataset was used in the CRAC paper~\cite{PSCR2013},
\item C.\ elegans (67.6M reads of 100\,bp) --- 
\url{http://ftp.sra.ebi.ac.uk/vol1/fastq/SRR065/SRR065390/}.
\end{itemize}
The command lines of the examined programs can be found in the PgSA package available at project homepage \url{http://sun.aei.polsl.pl/pgsa}.

\begin{table}
\centering
\caption{Dataset characteristics}
\label{table:datasets}
\begin{tabular}{lcccc}\toprule
Dataset & No.\ reads [M] & Read length & Alphabet size & PG length [MB] \\
\midrule
E.\ coli  & 11.5 & 151 & 5 & \q551.4 \\
GRCh37  & 42.4 & \q75 & 4 & \q567.9 \\
C.\ elegans & 67.6 & 100 & 5 & 1603.1 \\
\bottomrule
\end{tabular}
\end{table}

\begin{figure}[t]
\centering\includegraphics[width=\textwidth]{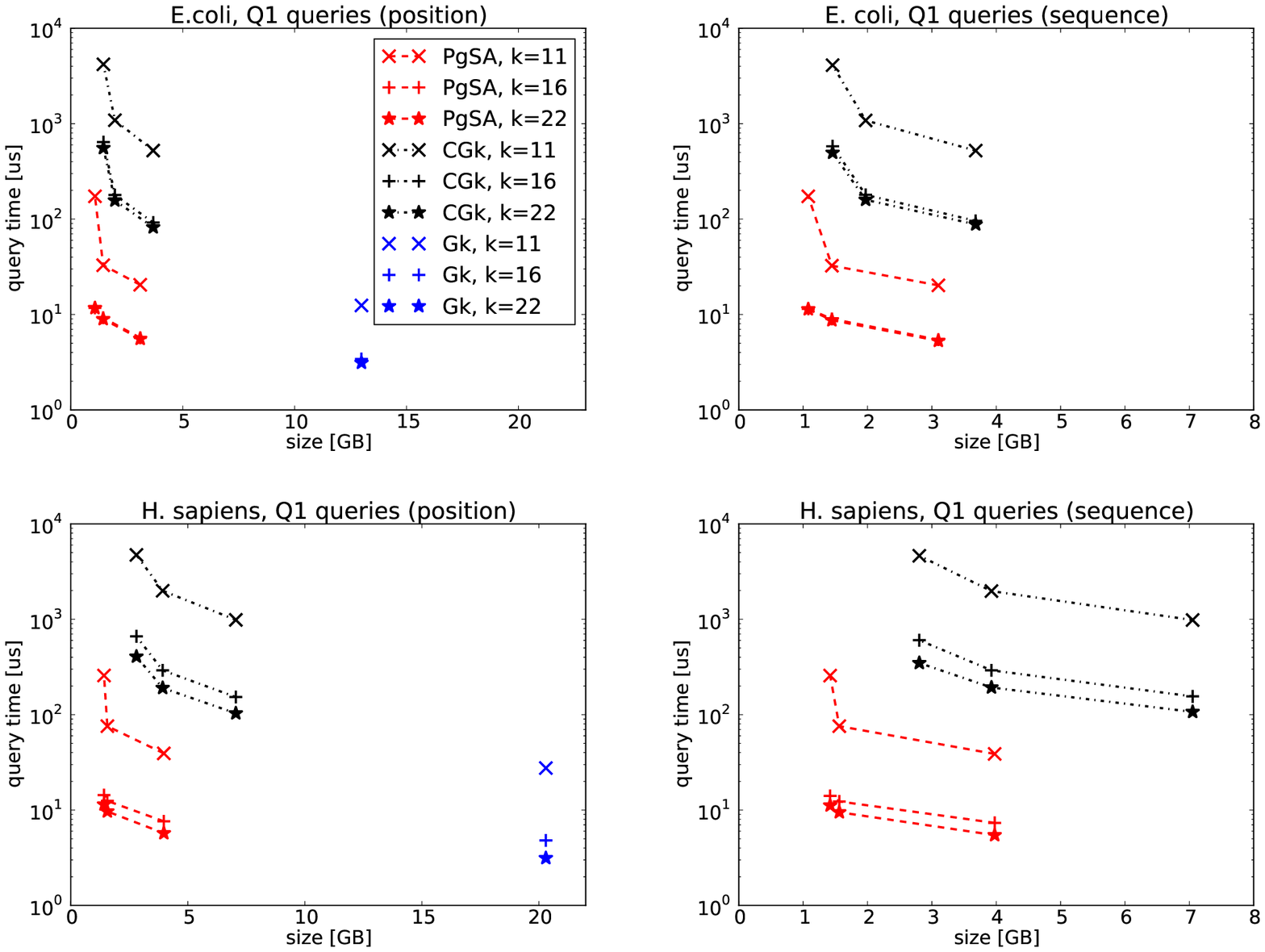}
\caption[Results]
{$Q1$ query results on E.\ coli (top row) and H.\ sapiens (bottom row) data. On the left figures the query is given as a position in the read list, while on the left ones as a string.}
\label{fig:q1}
\end{figure}

\begin{figure}[pt]
\centering\includegraphics[width=\textwidth]{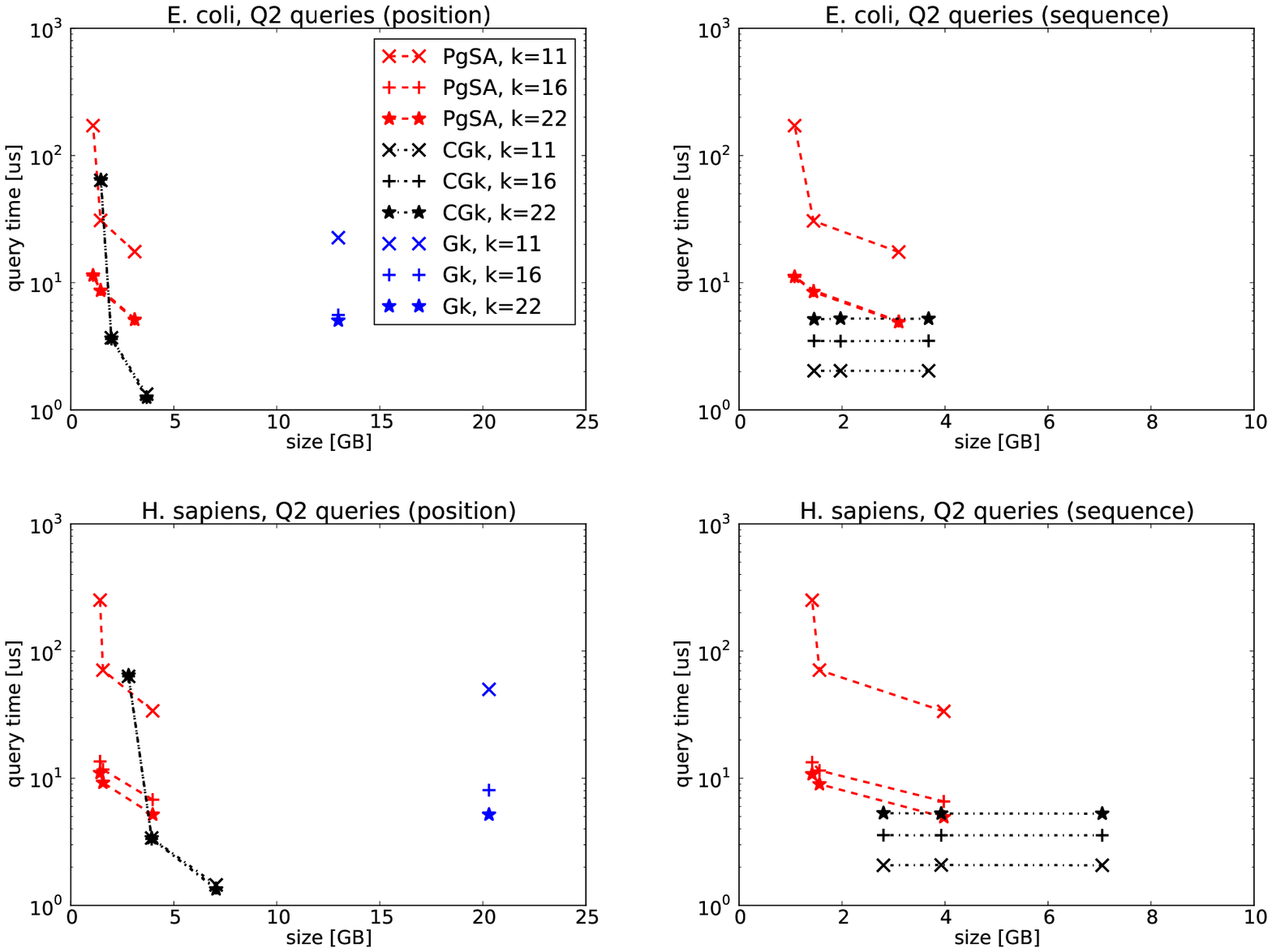}
\caption[Results]
{$Q2$ query results on E.\ coli (top row) and H.\ sapiens (bottom row) datasets. On the left figures the query is given as a position in the read list, while on the left ones as a string.}
\label{fig:q2}
\end{figure}

\begin{figure}[pt]
\centering\includegraphics[width=\textwidth]{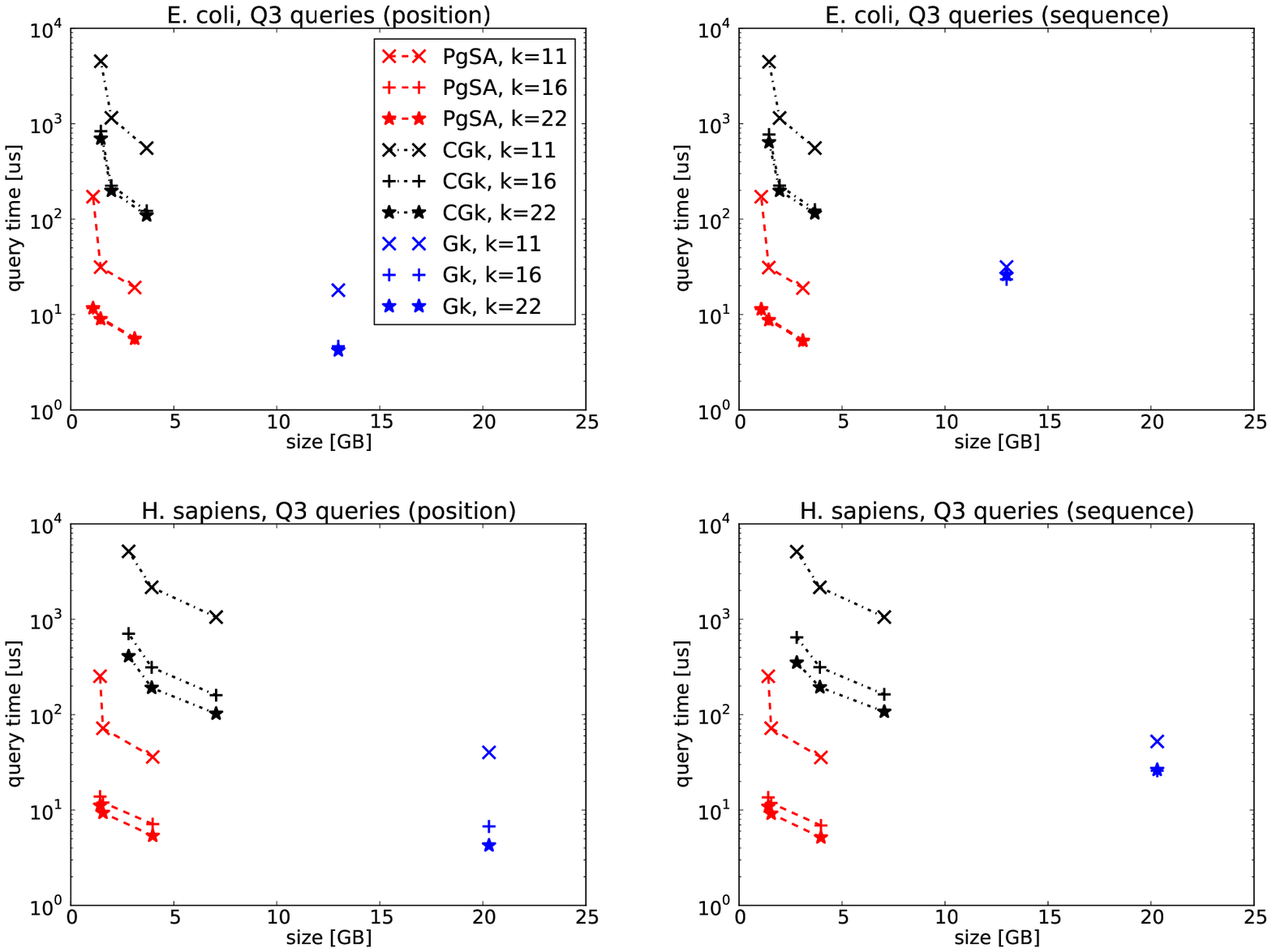}
\caption[Results]
{$Q3$ query results on E.\ coli (top row) and H.\ sapiens (bottom row) datasets. On the left figures the query is given as a position in the read list, while on the left ones as a string.}
\label{fig:q3}
\end{figure}

\begin{figure}[pt]
\centering\includegraphics[width=\textwidth]{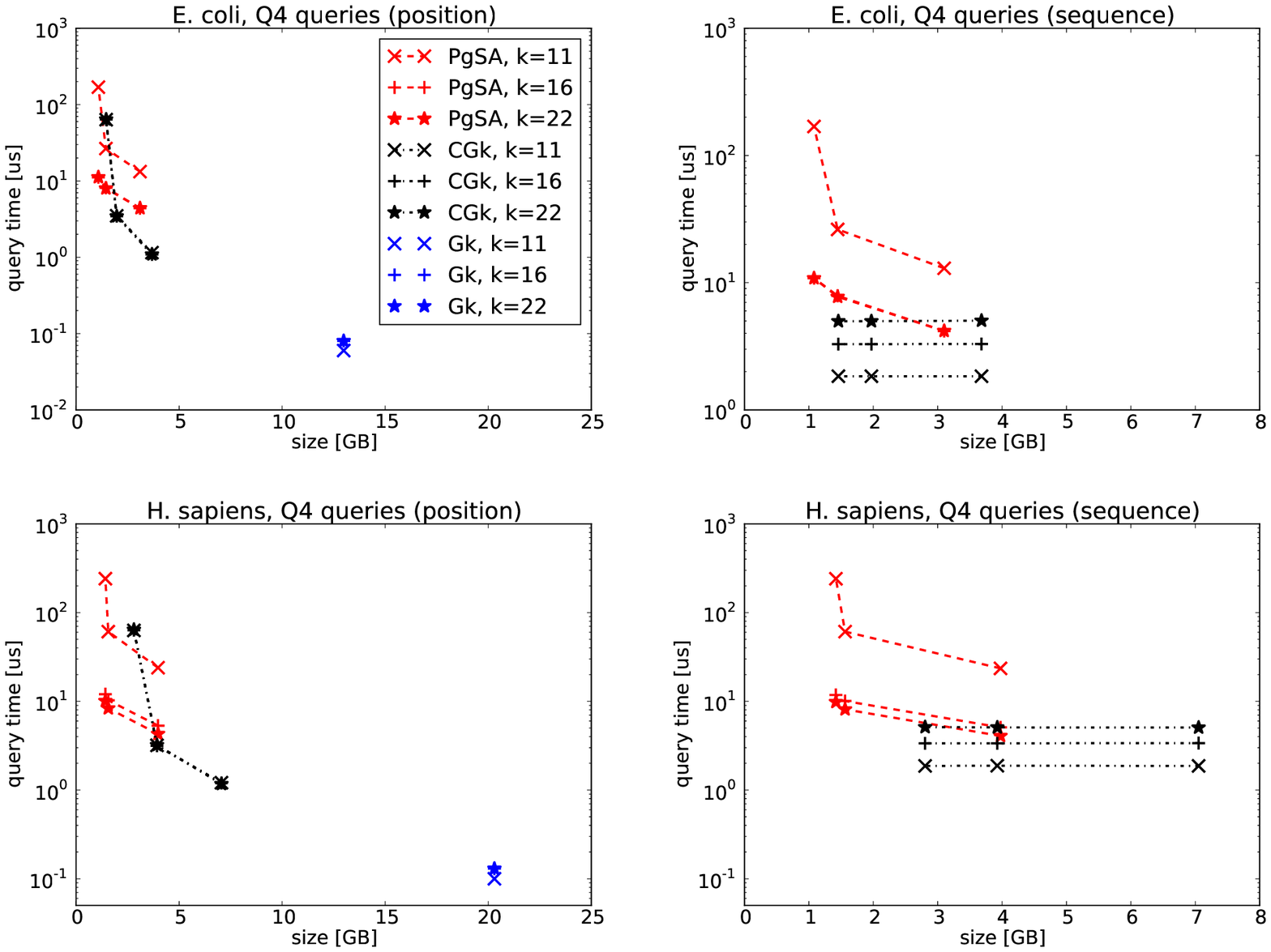}
\caption[Results]
{$Q4$ query results on E.\ coli (top row) and H.\ sapiens (bottom row) datasets. On the left figures the query is given as a position in the read list, while on the left ones as a string.}
\label{fig:q4}
\end{figure}

In the first experiments, we compare PgSA versus GkA (version 2.1.0) and CGkA 
(version cgka\_2013\_08\_21)
on two datasets, E.\ coli and GRCh37-75bp-simulated 
reads (Figs~\ref{fig:q1}, \ref{fig:q2}, \ref{fig:q3}, \ref{fig:q4}).
We can see that in $Q1$ and $Q3$ queries PgSA is by more than an order 
of magnitude faster than CGkA at comparable or better compression rate.
As expected, GkA is faster than CGkA (and sometimes faster, although not 
very significantly, than PgSA), yet requiring at least 3 times more space.
The speed relation is different for $Q2$ and $Q4$ queries.
Here CGkA defeats PgSA, sometimes by an order of magnitude.
In the $Q4$ query, given by position, GkA is a clear winner in speed.
We note that the tested (latest) GkA version (v2.1.0)
does not support $Q1$, $Q2$ and $Q4$ when the query is given as a sequence rather
than a position.
Overall, we believe that PgSA offers attractive space-time tradeoffs for most queries, 
and in contrast to its competitors it handles arbitrary values of $k$ 
(rather than a fixed one).
Additionally, we note that the latest GkA and CGkA versions do not support 
the $Q5$--$Q7$ queries.


\begin{figure}[pt]
\includegraphics[width=\textwidth]{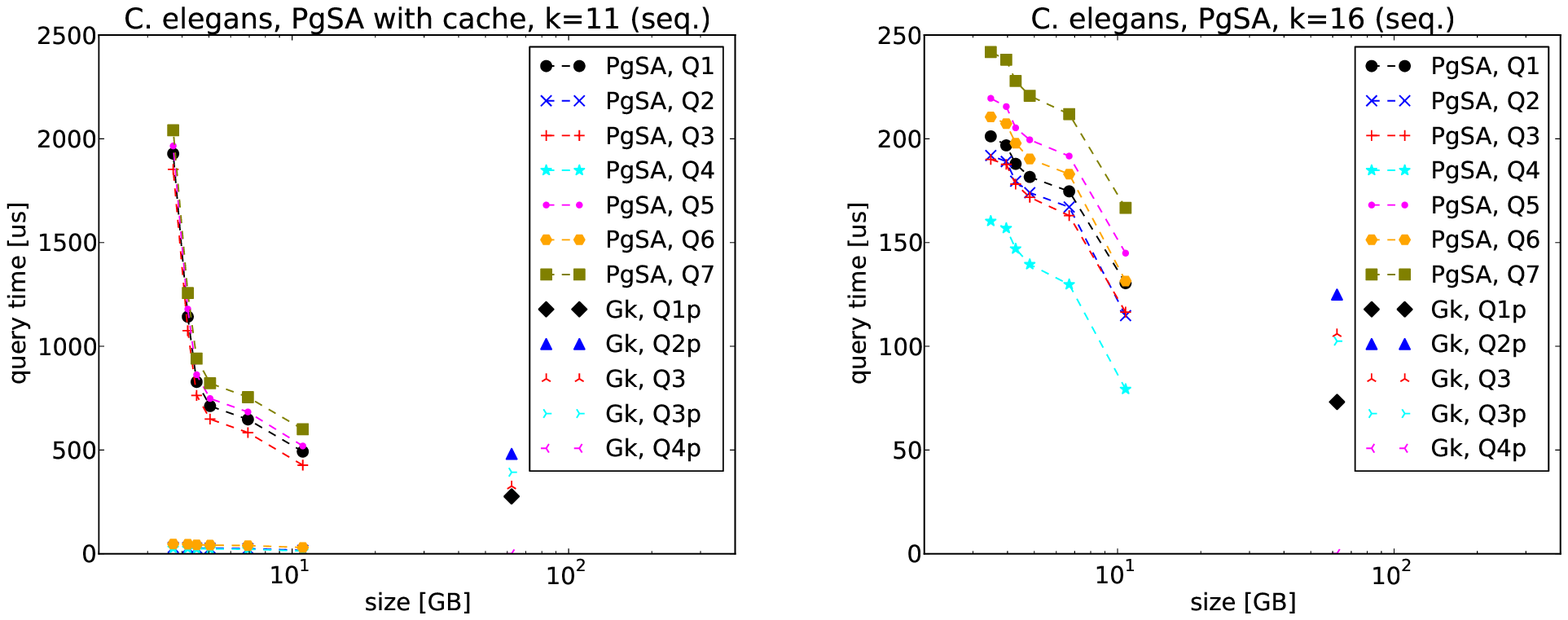}
\caption[Results]
{$Q1$--$Q7$ query results of PgSA and GkA on C.\ elegans dataset.
The letter `p' appended to some query names means that the query is
given as a position in the read list.}
\label{fig:celegans}
\end{figure}

In the next experiment we ran only PgSA and GkA on C.\ elegans dataset (Fig.~\ref{fig:celegans}).
We were not able to run CGkA on this dataset.
The PgSA lines on the 
figures are for the queries $Q1$--$Q7$ given as a sequence
(the time differences with regard to queries given as a position are up to 1 percent),
and the left and right figure corresponds to the query length $k=11$ and $k=16$, respectively.
Note that the results for the queries $Q2$, $Q4$, and $Q6$ are precomputed (cached) 
for $k = 11$.

In Tables~\ref{table:ecoli-space} and~\ref{table:celegans-space}
we detail out how much space is consumed by 
the components of the PgSA solution.

\begin{table}
\centering
\caption{E.\ coli dataset, space consumption. All sizes in megabytes.}
\label{table:ecoli-space}
\begin{tabular}{cccccc}\toprule
SA sparsity & $PG$ & $\mathcal{R}^\text{PG}$ & $SA^\text{PG}$ & $LUT$ & total\\
\midrule
1 & 551 & 149 & 2205 & 195 & 3101 \\
2 & 276 & 149 & 1378 & 195 & 1999 \\
3 & 184 & 149 &  919 & 195 & 1447 \\
4 & 276 & 149 &  689 & 195 & 1309 \\
5 & 221 & 149 &  662 & 195 & 1227 \\
6 & 184 & 149 &  551 & 195 & 1080 \\
\bottomrule
\end{tabular}
\end{table}

\begin{table}
\caption{C.\ elegans dataset, space consumption. All sizes in megabytes.}
\label{table:celegans-space}
\centering
\begin{tabular}{cccccc}\toprule
SA sparsity & $PG$ & $\mathcal{R}^\text{PG}$ & $SA^\text{PG}$ & $LUT$ & total\\
\midrule
1 & 1603 & 879 & 8016 & 195 & 10693 \\
2 & \q802 & 879 & 4809 & 195 & \q6685 \\
3 & \q534 & 879 &  3206 & 195 & \q4814 \\
4 & \q802 & 879 &  2405 & 195 & \q4280 \\
5 & \q641 & 879 &  2244 & 195 & \q3959 \\
6 & \q534 & 879 &  1870 & 195 & \q3479 \\
\bottomrule
\end{tabular}
\end{table}

\begin{table}
\caption{CRAC, $k = 22$. Times in minutes, sizes in gigabytes.}
\label{table:crac}
\centering
\begin{tabular}{lccccc}\toprule
Type & Build & Build+CRAC & Index & Max mem. & Max mem. \\
     & time  &  time      & size  & (build)  & (CRAC) \\
\midrule
PgSA $s=1$  &  50.5 &  410.7 &  \q3.98 &  \q8.16 &  \q6.06 \\
PgSA $s=4$ &  36.3 &  509.0 &  \q1.56 &  \q8.16 &  \q3.65 \\
PgSA $s=6$ &  34.9 &  572.3 &  \q1.42 &  \q8.16 &  \q3.51 \\
Gk          &  11.6 &  218.9 & 20.30 & 27.60 & 21.98 \\
\bottomrule
\end{tabular}
\end{table}

Finally, we checked how replacing GkA with PgSA affects the CRAC performance
(Table~\ref{table:crac}).
We used CRAC v1.3.2 (\url{http://crac.gforge.inria.fr}).
Unfortunately, the build time grows several times (and even including the CRAC 
processing time the difference is at least by factor 2), 
yet the memory requirements of the PgSA-based variant are significantly lower, 
which may be a crucial benefit if a low-end workstation is only available.

\section*{Discussion}
We proposed a new indexing structure for read collections.
The experiments proved that this structure is much more compact than the existing solutions, GkA and CGkA.
The running times of the 
counting
queries are worse than of the CGkA, 
but in the listing queries PgSA is usually faster.

Several aspects of the presented scheme can be improved.
We have noticed that using both direct and reverse-complemented 
reads in our construction of the pseudogenome reduces its size by about 15\%.
Still, this easy change for the construction is not equally easy to handle 
during the search, therefore the current implementation refrains from it.
Additionally, our recent progress with read compression~\cite{GDR2014} 
suggests to build the pseudogenome from large datasets on disk 
(disk-based SA construction algorithms also exist, see, e.g.,~\cite{BFO2013} 
and references therein).
Finally, the sparse suffix array may be replaced by a recent sparse index, 
SamSAMi (sampled suffix array with minimizers) \cite{GR2014}, with hopefully 
better performance.



%
%

\section*{Acknowledgments}
The Polish National Science Centre under the project DEC-2012/05/B/ST6/03148.
The infrastructure supported by POIG.02.03.01-24-099/13 grant: `GeCONiI---Upper
Silesian Center for Computational Science and Engineering'.


%
%
%



\end{document}